\documentclass[aps,prl,onecolumn,a4paper]{revtex4}
\usepackage[dvips]{graphicx}
\usepackage[dvips]{color}
\usepackage{subfigure}
\usepackage{epsfig}
\usepackage{dcolumn}
\usepackage{bm}
\usepackage{amsmath}
\usepackage{amssymb} 
\usepackage{hyperref}

\begin{document}

\title{The Adam-Gibbs relation for glass-forming liquids in 2, 3 and 4 dimensions}
\author{Shiladitya Sengupta$^{1}$, Smarajit Karmakar$^{2}$, Chandan  Dasgupta$^{3}$, Srikanth  Sastry$^{1,4}$} 
\affiliation{
$^{1}$ Theoretical Sciences Unit, Jawaharlal Nehru Centre for Advanced Scientific Research, 
Jakkur Campus, Bangalore 560 064, India.\\ 
$^{2}$ Departimento di Fisica, Universita di Roma `` La Sapienza ", Piazzale Aldo Moro 2, 00185,
Roma, Italy.\\
$^{3}$ Centre for Condensed Matter Theory, Department of Physics, Indian Institute of
  Science, Bangalore, 560012, India.\\
$^{4}$ TIFR Centre for Interdisciplinary Sciences, 21 Brundavan Colony,
Narsingi,Hyderabad 500075, India.}
\date{\today}

\begin{abstract}
The Adam-Gibbs relation between relaxation times and the
configurational entropy has been tested extensively for glass formers
using experimental data and computer simulation results.  Although the
form of the relation contains no dependence on the spatial
dimensionality in the original formulation, subsequent derivations of the
Adam-Gibbs relation allow for such a possibility. We test the
Adam-Gibbs relation in 2, 3, and 4 spatial dimensions using computer
simulations of model glass formers. We find that the relation is valid
in 3 and 4 dimensions. But in 2 dimensions, the relation does not
hold, and interestingly, no single alternate relation describes the
results for the different model systems we study.
\end{abstract}

\pacs{xx}

\maketitle

A central theme in the study of glass forming liquids is a
satisfactory understanding of the behaviour of relaxation
times as the glass transition is approached, showing temperature dependence
typically stronger  than the Arrhenius law: $\tau(T) = \tau(\infty)
\exp[\frac{E_{0}}{k_{B}T}] \label{eqn:Arrh}$ observed at high
temperatures \cite{pap:Ediger-review, pap:angell-review-relaxation,pap:deb-stil}. 
The Adam-Gibbs (AG) relation \cite{pap:AG-2}, which  is of central importance in 
glass forming liquids, explains the behaviour of the  relaxation time {\it i.e.}  dynamics 
in  terms of the configurational entropy  {\it i.e.}  thermodynamics. It also  forms the
basis  of more sophisticated theories of glass transition connecting  dynamics to thermodynamics 
{\it e.g.} the random first order transition theory (RFOT) \cite{pap:RFOT-Kirkpatrick-etal,parisi-mezard,mosaic,rfotrev,pap:AG-RFOT-Bouchaud-Biroli}.
In  these theories  the  spatial  dimensionality  (D)  appears explicitly  in  the
relationship between  dynamics and thermodynamics.   However, the large number  of experimental
and numerical studies  wherein the AG relation has  been tested  \cite{pap:AG-Greet-Turnbull, pap:AG-Richert-Angell-V, pap:AG-Speedy, pap:AG-Sastry, pap:AG-Johari, pap:AG-Scala-etal, pap:AG-Mossa-etal, pap:AG-Rolandetal}
have all been  in 3 dimensions. The present study aims to critically examine whether the AG relation is
valid in different spatial dimensions or gets generalized in a D dependent way.

The AG relation is  based on the picture that relaxation
in glass forming liquids occurs through the collective rearrangement
of ``cooperatively rearranging regions'' (CRR). The CRRs define a
minimum size of groups of rearranging particles (atoms, molecules {\it
  etc.} depending on the nature of the glass former) such that smaller
groups of particles are incapable of rearrangement independently of
their surroundings. Adam and Gibbs argued that the configurational
entropy  (the entropy associated with the multiplicity
of distinct arrangements of particles, obtained by subtracting a
``vibrational'' component from the total entropy) per particle $S_{c}(T)$ of a liquid varies
inversely as the size of the CRR, $z(T)$, since the configurational
entropy per CRR, $S^{*}$, is roughly independent of temperature:
$S_{c}(T) = \frac{S^{*}}{z(T)}$. The further assumption that the free
energy barrier for a rearrangement is proportional to the size of the
CRR ($\Delta G = z \delta \mu, \delta \mu=$ chemical potential barrier
per particle) results in the Adam-Gibbs relation:
\begin{eqnarray}
\tau(T) &=&  \tau(\infty)  \exp\left(\frac{S^{*} k_{B}^{-1} \delta \mu}{TS_{c}(T)} \right) \nonumber\\
            &=&  \tau(\infty)  \exp\left(\frac{C}{TS_{c}}\right)
            \label{eqn:AG}
\end{eqnarray}
The above relation is obtained independent of reference to the spatial
dimensionality of the system and is hence expected to be same in all
spatial dimensions. 

A rationalization of the AG relation, based on more detailed
considerations of possible activated relaxation mechanisms, is offered
by the RFOT, which
has recently been discussed by many authors specifically in the
context of growing length scales associated with the glass transition
\cite{silvio_kac,dalle-ferrier,pap:RFOTexpo-Capaccioli-etal,birolietal-natphys,pap:RFOTexpo-Cammarota-etal,pap:AG-Karmakar,thesis:Karmakar}. 


In the ``mosaic'' picture \cite{pap:RFOT-Kirkpatrick-etal,mosaic}, the
liquid is divided into metastable regions of characteristic size
$\xi(T)$. The transition from one metastable state to others is
hindered by the cost of surface free energy ( $\Delta G_s \propto Y\xi^{\theta}$,
where $\theta \leq D$, where $D$ is the spatial dimension and $Y$ is the
surface tension) and driven by the possibility of sampling an
exponentially large number of other metastable minima (free energy
gain $\Delta G_b \propto TS_{c}\xi^{D}$).  There is a characteristic cross-over
length $\xi^{*}(T) \propto
\left(\frac{Y}{TS_{c}}\right)^{\frac{1}{D-\theta}}$ above which a
liquid samples all metastable states and below which the liquid is
trapped in one of the metastable states.  The characteristic length
$\xi(T)^{*}$ diverges as $S_{C}(T) \rightarrow 0$. Assuming in general
that the free energy barrier to relaxation varies (following the
notation of \cite{pap:AG-RFOT-Bouchaud-Biroli}) as $\Delta G (T)
\propto \xi(T)^{\psi}$, one obtains
\begin{eqnarray}
 \tau(T) &=& \tau(\infty) \exp\left[\frac{A}{\left({TS_{c}} \right)^{\frac{\psi}{D-\theta}}}\right]           
\label{eqn:AG-RFOT}
\end{eqnarray}
where $A$ has weak $T$ dependence, and $\psi = \theta$ if $\Delta G
(T)$ is calculated as the free energy barrier obtained from the
surface and bulk contributions above ($\Delta G = \Delta G_b + \Delta
G_s$). The mosaic picture thus contains an explicit dependence on the
spatial dimesion, and one recovers the Adam Gibbs relation only when
$\frac{\psi}{D-\theta}=1$.  The original AG formulation is equivalent
to assuming $\psi = D$, $\theta = 0$, whereas it was argued in
\cite{pap:RFOT-Kirkpatrick-etal} that $\theta = \psi = D/2$, with the
AG relation as the result in both cases. In the latter case
\cite{pap:RFOT-Kirkpatrick-etal}, a dimension independent AG relation
is predicted in spite of arguments that treat spatial dimensions
explicitly. 

However, attempts to estimate the exponents either numerically or from
experimental data are not conclusive. By direct simulation in a model
liquid ($D = 3$), Cammarota {\it et al.} obtained $\theta=2$, $\psi=1$
\cite{pap:RFOTexpo-Cammarota-etal} which, although consistent with the
AG relation, differs from both the AG and RFOT values. The simulation
study by Karmakar {\it et al.} \cite{pap:AG-Karmakar,thesis:Karmakar}
obtained $\theta \sim 2.3$ and hence $\psi \sim 0.7$ using the
condition $\frac{\psi}{D-\theta} = 1$. An extensive study by
Capaccioli {\it et al.} \cite{pap:RFOTexpo-Capaccioli-etal} of
experimental data for 45 glass-forming liquids show a lot of variation
in both $\theta$ and $\psi$, with best fit estimates in the range of
$\theta \sim 2 - 2.15$, and $\psi \sim 0.85 - 1$. These two studies
are consistent in suggesting values of $\theta > 2$, $\psi < 1$, but
both use measures of dynamical heterogeneity to extract a length
scale. Since {\it a priori} one must expect the mosaic length to be
distinct from the heterogeneity length scale, the implications of
these estimates are not clear.

In view of the considerations above, the possibility of an explicit
dependence of the AG relation on the spatial dimension (and within the
RFOT framework, the values of the exponents $\theta$ and $\psi$) merits
investigation.  In this letter, we address the question of the
dependence on spatial dimension of the AG relation by studying model
liquids in $2$, $3$ and $4$ dimensions, and evaluating the relationship between 
relaxation times and the configurational entropy. The evaluation of the RFOT exponents 
$\theta$ and $\psi$, while relevant to our investigation,
requires the calculation of the mosaic lengthscale. Ways of obtaining lengthscales
relevant to glassy dynamics that have been discussed in the literature (see e.g., 
\cite{birolietal-natphys,pap:xiPTS-Hockyetal,pap:AG-Karmakar,pap:atrep-karmakar})
tend to be computationally very demanding \cite{pap:Karmakar-PRL10}. We have therefore
not attempted to evaluate the RFOT exponents in this work.

 One of the well-known systems in which the AG
relation is shown to be valid \cite{pap:AG-Sastry,
  pap:AG-Karmakar,Shila} is Kob-Andersen model (KA) \cite{pap:KA},
which is an $80:20$ binary mixture of particles interacting with a
Lennard-Jones (LJ) potential, with LJ parameters
$\epsilon_{AB}/\epsilon_{AA}=1.5$, $\epsilon_{BB}/\epsilon_{AA}=0.5$,
$\sigma_{AB}/\sigma_{AA}=0.80$, $\sigma_{BB}/\sigma_{AA}=0.88$.  We
truncate the interaction potential at $2.5 \sigma_{\alpha\beta}$
(details as in \cite{pap:AG-Sastry}). Units of length, energy and time
scales are $\sigma_{AA},\epsilon_{AA}$ and
$\sqrt{\frac{\sigma_{AA}^{2}m_{AA}}{\epsilon_{AA}}} $ respectively. We
have studied the KA model in two (2D), three (3D) and four (4D)
spatial dimensions. In addition in two dimensions we have studied the
following liquids: (1) KA model at 65:35 composition \cite{pap:KA-2D}
which is denoted as the modified KA (MKA) model and (2) 50:50 binary
mixture of repulsive soft spheres ($R10$ model, $V(r) \sim r^{-10}$)
(see \cite{pap:r10defnSK} for details of the model. Like the KA
models, we use reduced units defined in terms of the energy scale of
the model and the size of the large particles). We have performed
molecular dynamics simulations in the canonical (NVT) ensemble, using
the constant temperature algorithm of Brown and Clarke
\cite{pap:BC}. Simulations were done at a fixed number density $\rho$
($\rho =1.2$ for the KA model in two and three dimensions, $\rho =
1.6$ in four dimensions, $\rho = 1.2$ for the MKA model and $\rho =
0.85$ for the R10 model). Integration time steps were in the range
$\in [0.001, 0.006]$ depending on $T$. Run lengths at each
$T$ were in excess of $100 \tau_{\alpha}$ (the relaxation time
$\tau_{\alpha}$ is defined below) ({\it e.g.}, at the lowest
$T$ in 4D the total runlength is $\sim 6 \times 10^{8}$ MD
steps). The range of relaxation times accessed is upto
$\mathcal{O}{(10^{4})}$ in three and four dimensions and upto
$\mathcal{O}{(10^{5})}$ in two dimensions. Typically, $3$ ($1 - 3$ in 4D, $3 - 5$ in 3D and 2D)
independent runs are performed at each $T$. In order to assess
the influence of system size, we show results for different system
sizes in $2$ and $3$ dimensions (indicated in appropriate places)
although we do not discuss system size effects in detail here.


As a measure of dynamics, we have studied a two-point time correlation
function, the overlap function $q(t)$ and its fluctuation related to
dynamic susceptibility ($\chi_{4}(t)$)
\cite{pap:4pt-CD,pap:Ovlap-Glotzer-etal,pap:Ovlap-Donati-etal,pap:AG-Karmakar}:

\begin{eqnarray}
<q(t)> &=& <\int d\vec{r}\rho(\vec{r},t_{0})\rho(\vec{r},t+t_{0})>\nonumber\\
&\sim& <\sum_{i=1}^{N} w(|\vec{r}_{i}(t_{0}) - \vec{r}_{i}(t_{0}+t)|)>\nonumber\\
\chi_{4}(t) &=&{1\over N} \langle q(t)^{2} \rangle - \langle q(t) \rangle^{2}\nonumber
\end{eqnarray}
where $\rho(\vec{r},t_{0})$ is the space and time dependent particle
density, $w(r)$ is the window function: $w(r) = 1, r \leq a \mbox{ and zero otherwise}$ ($a=0.3$ in 3D)
and averages are obtained over initial
times $t_{0}$ as well as independent samples. Relaxation times are
estimated both from the condition $q(\tau)/N = 1/e$ ($\tau_\alpha$)
and the characteristic time where $\chi_{4}(t)$ is maximum
($\tau_4$). As we find these two times to be proportional to each
other \cite{pap:AG-Karmakar}, we report only $\tau_\alpha$.

The configurational entropy ($S_{c}$) per particle, the measure of the
number of distinct local energy minima, is calculated
\cite{pap:Sc-Sastry} by subtracting from the total entropy of the
system the ``vibrational'' component:
\begin{equation}
S_{c͑}(T) =  S_{total}(T) - S_{vib} (T)\nonumber
\end{equation}
The total entropy of the liquid is obtained via thermodynamic
integration \cite{pap:Sc-Sastry}. The vibrational entropy is calculated by
making a harmonic approximation to the potential energy about
local energy minima (termed ``inherent strcture''s)  \cite{pap:Sc-Sastry,pap:Sc-Sastry-JPCM,pap:PEL-Sciortino,pap:PEL-Heuer}.

\begin{figure}[h!]
\begin{center}
\subfigure[]{
\includegraphics[width=5.5cm, height=4cm]{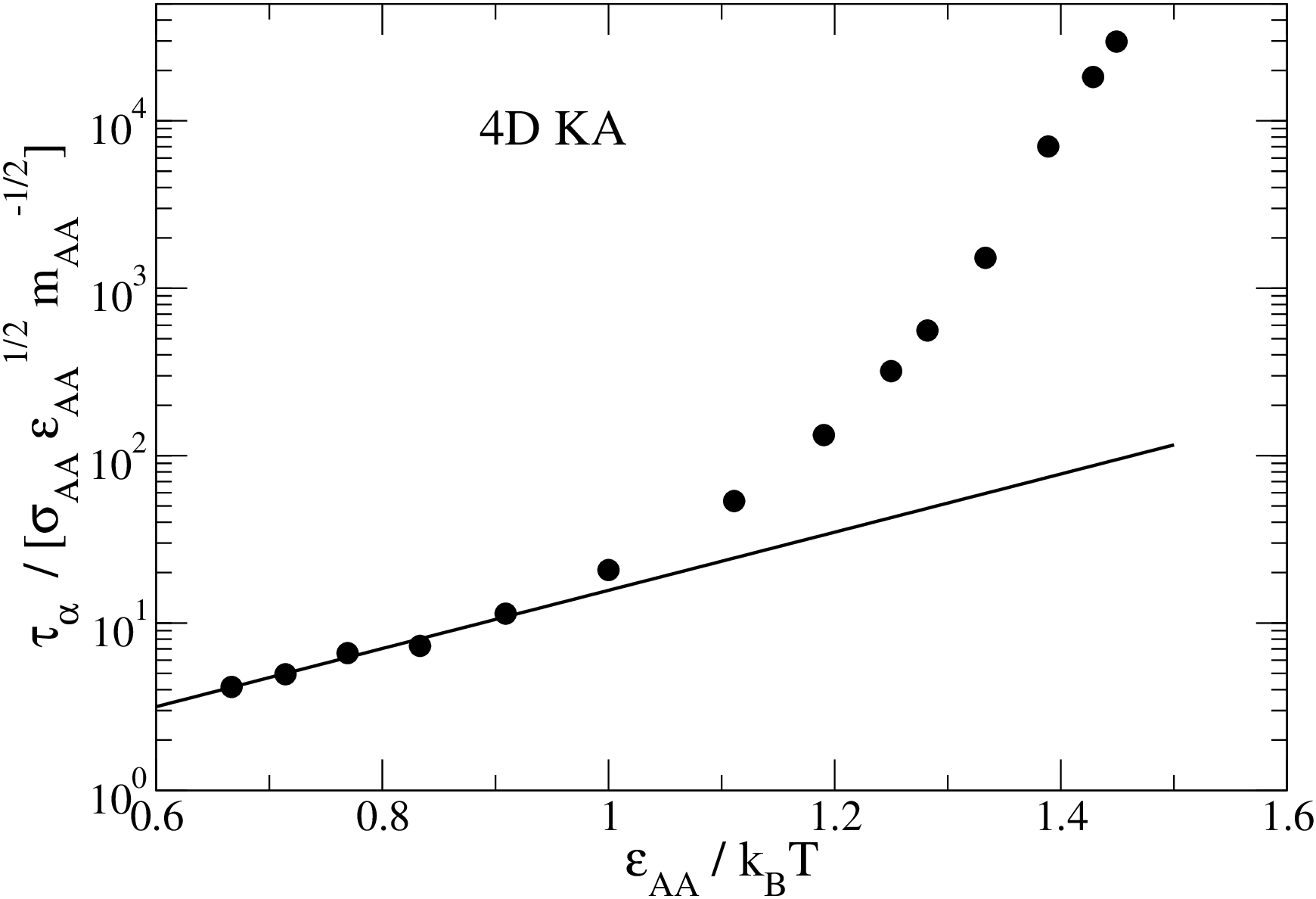}\label{fig:4D-Arrh-tau}
}\\\vspace{3mm}
\subfigure[]{
\includegraphics[width=5.5cm, height=4cm]{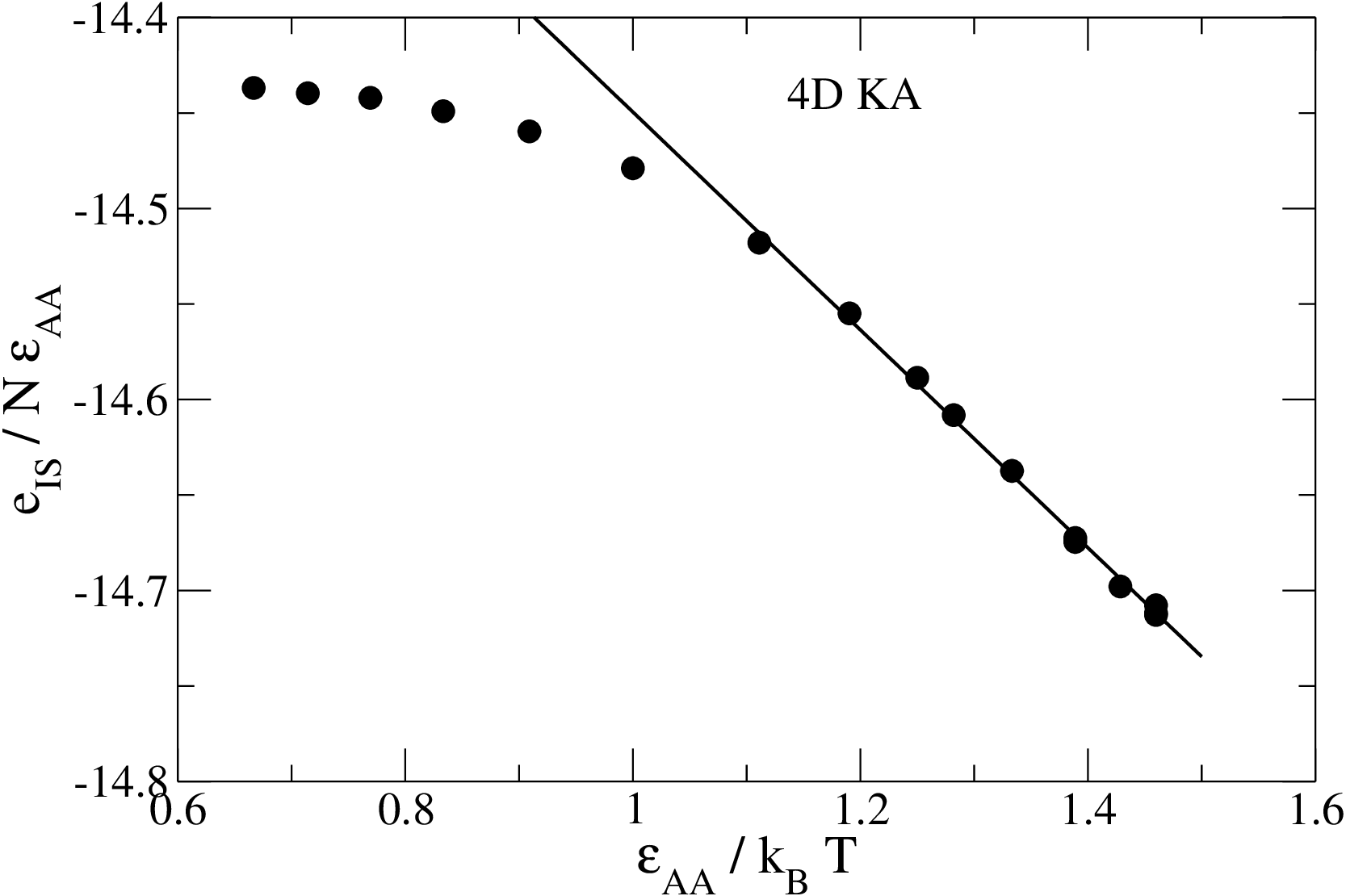}\label{fig:4D-Arrh-eIS}
}\\\vspace{2mm}
\subfigure[]{
\includegraphics[width=5.5cm, height=4cm]{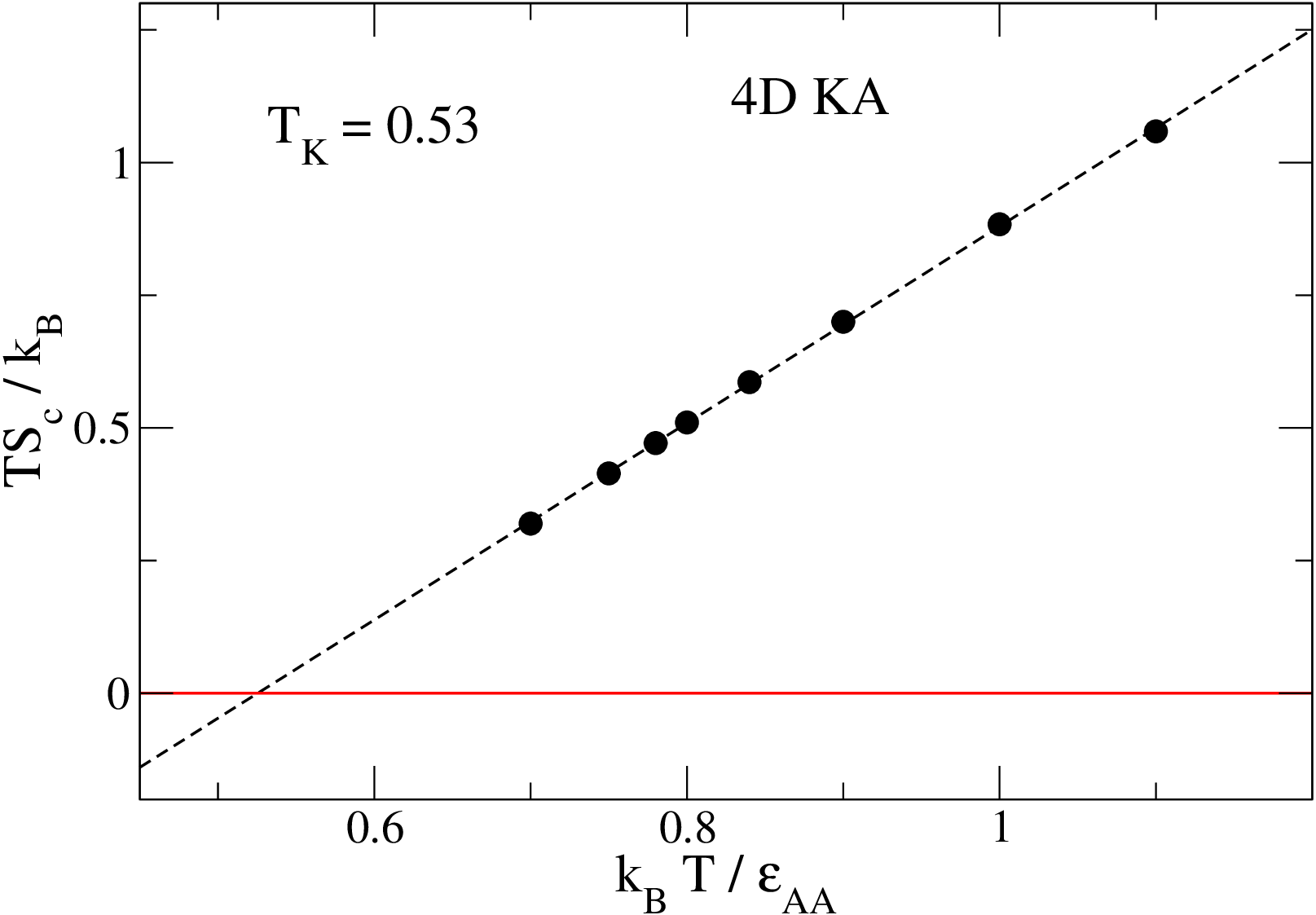} \label{fig:4D-TK}
}
\caption{\ref{fig:4D-Arrh-tau}: Temperature dependence of the
  relaxation time $\tau_\alpha$ of the 4D KA model displaying a
  cross-over from Arrhenius to non-Arrhenius behaviour at $T_{onset}
  \sim 1$. The solid line is an
  Arrhenius fit above $T_{onset}$. \ref{fig:4D-Arrh-eIS} shows the average inherent structure
  energy {\it vs.} $T$ displaying a $1/T$ dependence below
  $T_{onset}$.  \ref{fig:4D-TK}: Kauzmann
  temperature for the 4D KA model, $T_{K}=0.53$, is obtained from
  the condition $T_KS_{c}(T_{K})=0$. } \label{fig:4D-Arrh}
\end{center}
\end{figure}

We evaluate the configurational entropy below the onset temperature
($T_{onset}$) across which a cross-over from Arrhenius to
non-Arrhenius behaviour of the relaxation time occurs
\cite{pap:sastry-deb-stil,pap:AG-Sastry,pap:onset-sastry}. In
Fig. \ref{fig:4D-Arrh-tau} we show this cross-over in the 4D KA
model. Fig. \ref{fig:4D-Arrh-eIS} shows the temperature variation of
the inherent structure energy which exhibits a $1/T$ temperature
dependence below $T_{onset}$ \cite{pap:AG-Sastry}. In Fig. \ref{fig:4D-TK} we show the
$T$ dependence of the configurational entropy for the 4D KA model. As
$T$ is lowered $T S_{c}$ goes to zero linearly with
$T$. We estimate the Kauzmann temperature ($T_{K}$) at which
the extrapolated configurational entropy vanishes to be $T_{K}=0.53$.

\begin{figure}[h!]
\begin{center}
\subfigure[]{
\includegraphics[width=6.5cm, height=5cm]{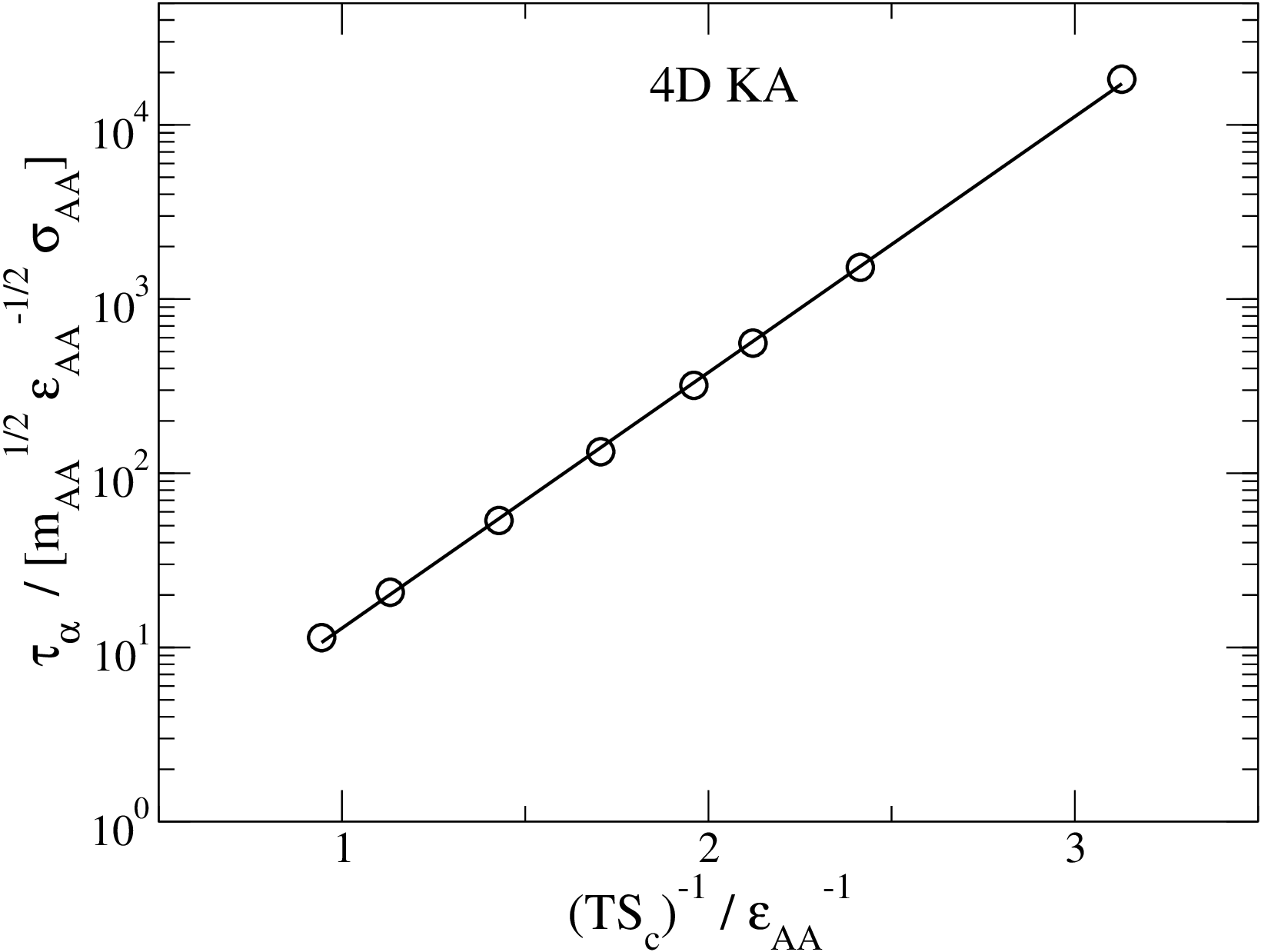}\label{fig:AG-4DKA}
}\\\vspace{2mm}
\subfigure[]{
\includegraphics[width=6.5cm, height=5cm]{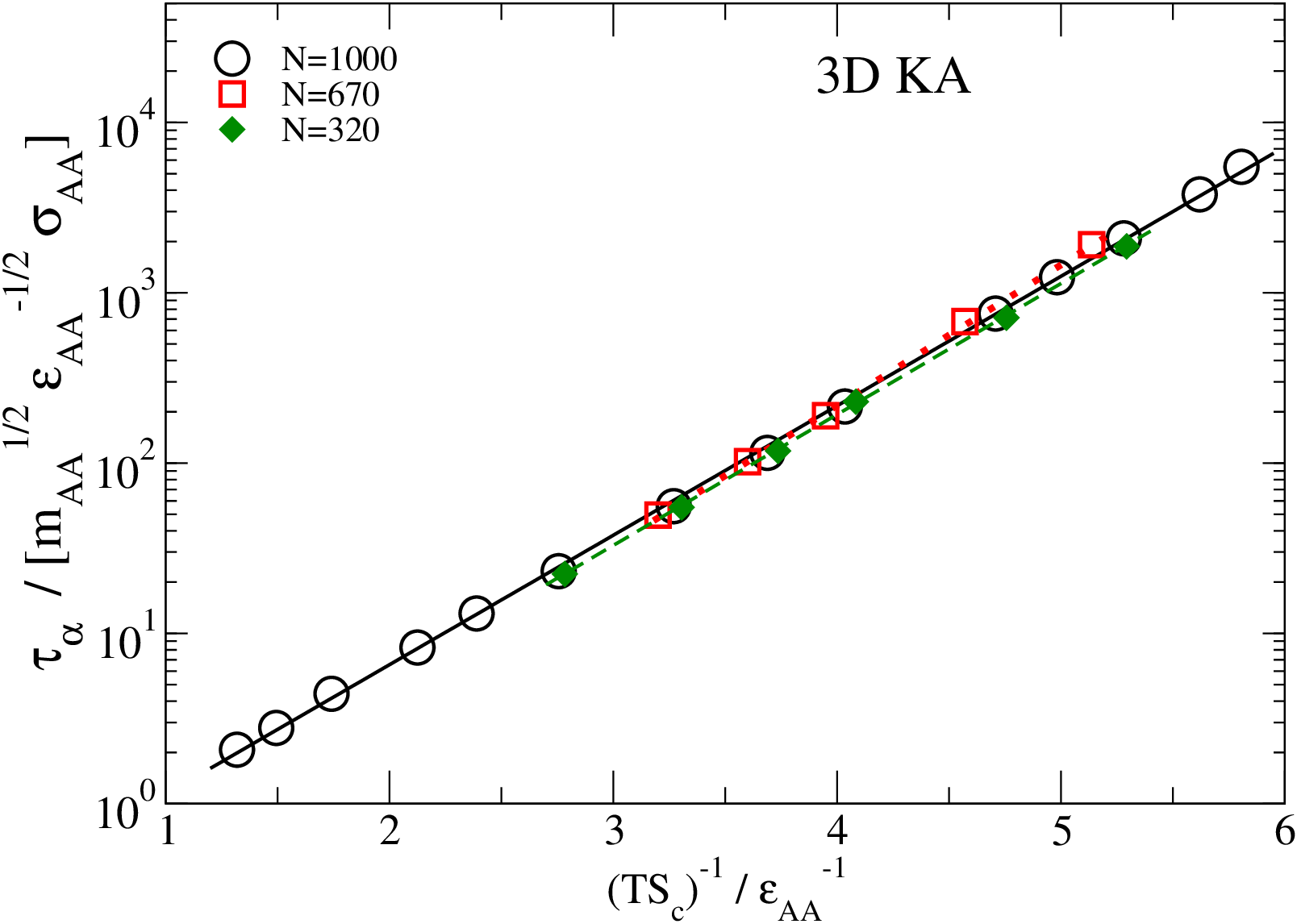}\label{fig:AG-3DKA}
}\\
\subfigure[]{
\includegraphics[width=6.5cm, height=5cm]{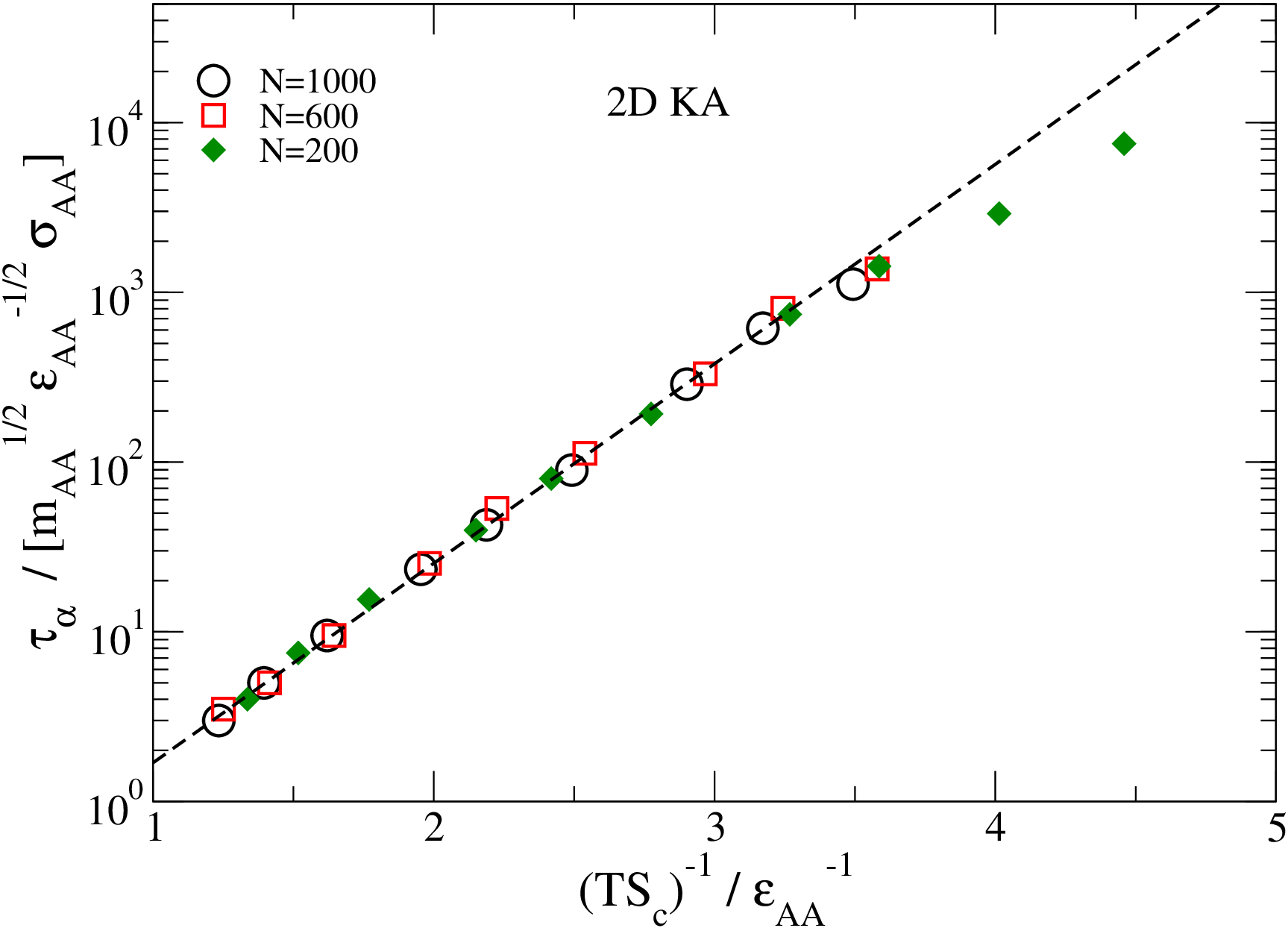}\label{fig:AG-2DKA}
}
\caption{\label{fig:AGKA} Adam-Gibbs plots ($\tau_\alpha$ {\it vs.}
  $(TS_c)^{-1}$) for the KA model in 4, 3, and 2 dimensions. The AG
  relation holds in three and four dimensions but deviations are observed
in two dimensions at low $T$ and small system sizes. The system size in 4D
  is $N=1500$. Lines are fits to the AG relation (Eq. \ref{eqn:AG}) (for
  the largest system size in the case of 2D).}
\end{center}
\end{figure}

\begin{figure}[h!]
\begin{center}

\subfigure[]{
\includegraphics[width=6.5cm, height=5cm]{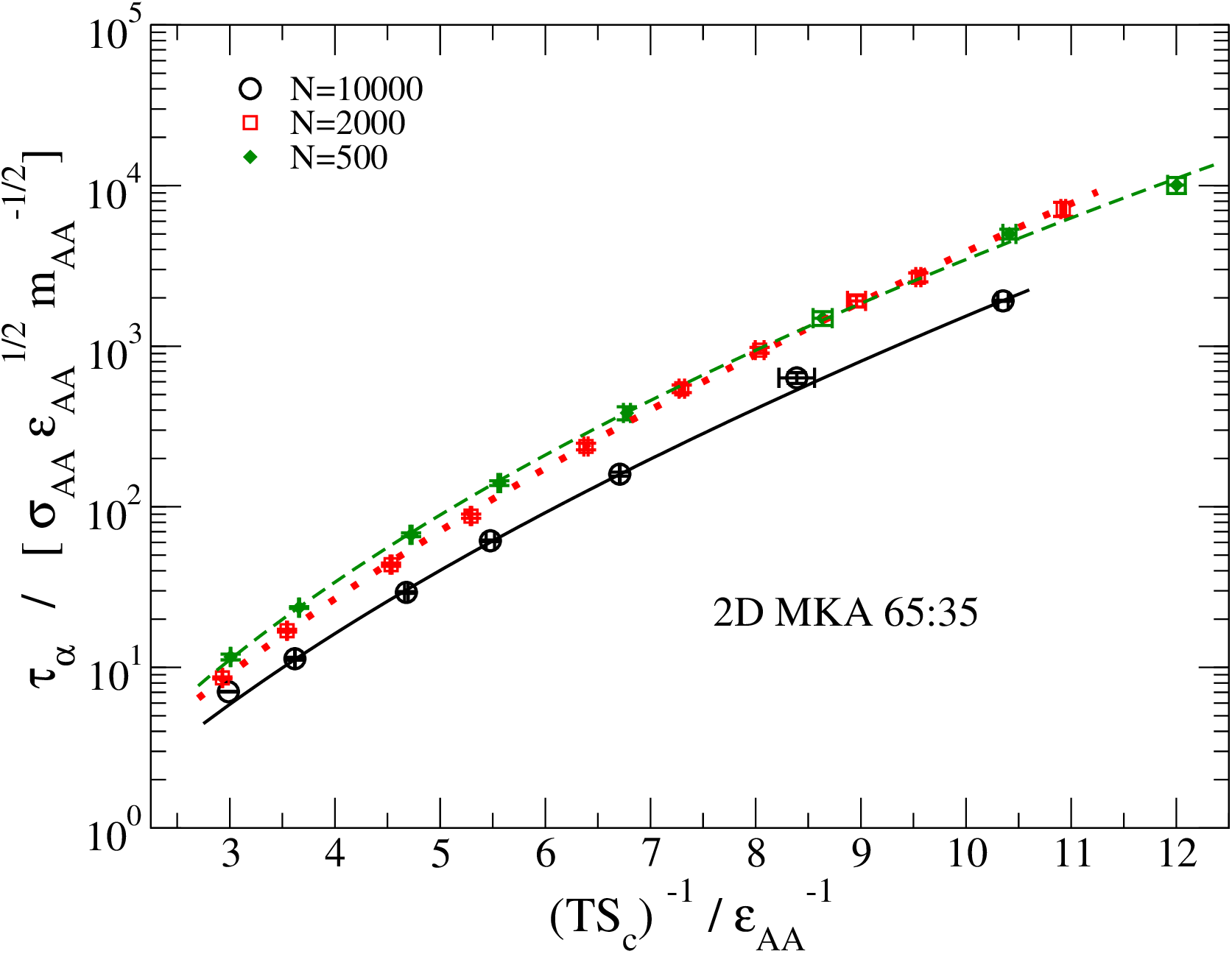}\label{fig:AG-2DMKA}
}\\
\subfigure[]{
\includegraphics[width=6.5cm, height=5cm]{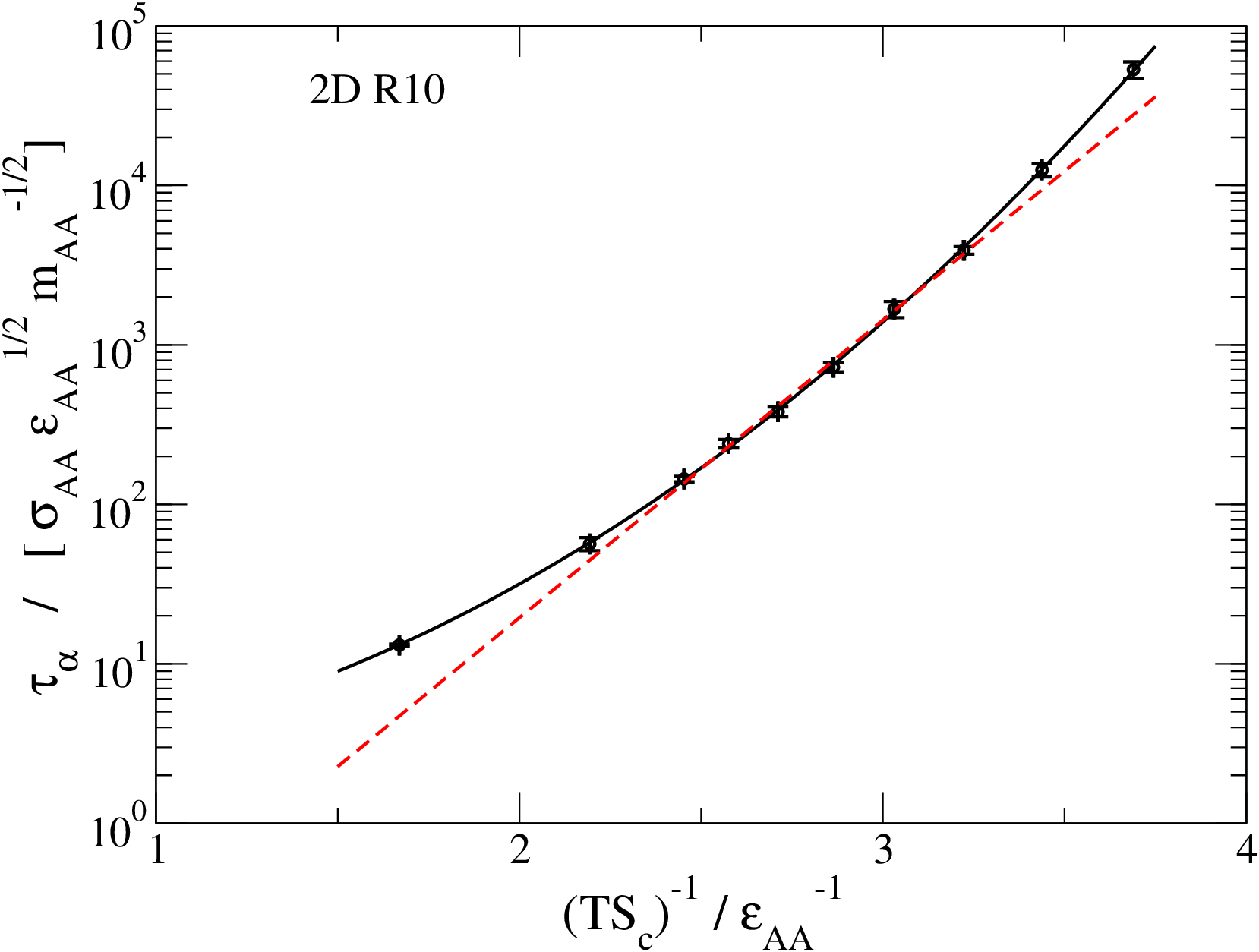}\label{fig:AG-2DR10}
}
\caption{Adam-Gibbs plot for the MKA and $R10$ models. Systematic
  deviations from the AG relation are seen for both models, but with
  opposite deviations from linearity. For the MKA model, the lines are fits to the
  generalized AG relation of the form $\tau_{\alpha} = \tau_0
  \exp\left((\frac{C}{TS_{c}})^{\alpha}\right)$ (see also
  Eq. \ref{eqn:AG-RFOT})[$\alpha=0.43$ for $N=500$
  (dashed), $\alpha=0.60$ for $N=2000$ (dotted line) and $\alpha=0.67$ for $N=10000$ (solid line)]. 
  For the $R10$ model, the dashed line is a
  fit to the AG relation and the solid line is a fit to the generalized AG relation with $\alpha=2.1$.
  The system size is $N=2048$ for the $R10$ model.
  }
\end{center}
\end{figure}

In Fig. \ref{fig:AGKA} we show the AG plot for the KA model in
different spatial dimensions. We see that in 4D, the AG relation is
valid over several orders of magnitude of change in $\tau_\alpha$. As
reported in earlier works, the AG relation is also valid in 3D, and
the data presented in Fig. \ref{fig:AG-3DKA} indicates a mild system
size dependence in the slopes $C$ (Eq. \ref{eqn:AG}). However, in two
dimensions [Fig: \ref{fig:AG-2DKA}], we see deviations from the AG
relation for low temperatures and small system sizes. It has been
reported recently that the 2D KA model is prone to orientational ordering
\cite{pap:KA-2D}, which we confirm (data not shown). It is therefore
not clear to what extent these observed deviations are significant.
In order to asertain the dependence of $\tau_\alpha$ on $S_c$ without
the presence of orientational ordering, we study two other previously
studied 2D models which do not exhibit significant ordering, namely
the KA model with composition $65:35$ (MKA) \cite{pap:KA-2D} and the
$50:50$ binary mixture of repulsive soft spheres ($R10$ model, $V(r)
\sim r^{-10}$) \cite{pap:r10defnSK}. Figs \ref{fig:AG-2DMKA}
and \ref{fig:AG-2DR10} show
that for these systems, $\tau_\alpha$ does not obey the AG
relation. 

We note that the nature of deviation is different in these two cases.
Fitting $\tau_\alpha$ to the generalized AG form $\tau = \tau_0
\exp\left((\frac{C}{TS_{c}})^{\alpha}\right)$, we find $\alpha=0.43$
for $N=500$, $\alpha=0.60$ for $N=2000$ and $\alpha=0.67$ for $N=10000$ ({\it i. e.} $\alpha < 1$)
for the MKA model, whereas for the $R10$ model we find $\alpha=2.1$
({\it i. e.} $\alpha > 1$). 

The MKA model has both attractive and repulsive interactions, whereas
the R10 model has purely repulsive interactions.  It has recently been
suggested \cite{pap:atrep-karmakar} that the exponent $\psi$ in the
relation $\tau(T) = \tau(\infty) \exp\left[\frac{A \xi^\psi}{k_B
    T}\right]$ (see discussion preceding Eq. \ref{eqn:AG-RFOT})
depends on the nature of the interactions. This possibility suggests
an explanation for the results we observe which is under
investigation. However, we note that in three dimensions, the AG relation
has been verified in both attractive models (e.g. Kob-Andersen
\cite{pap:AG-Sastry}, Lewis-Wahnstr\"{o}m OTP \cite{pap:AG-Mossa-etal}
and Dzugutov liquid \cite{pap:AG-Djugutovliq}) and repulsive models
(e.g. repulsive soft spheres \cite{pap:AG-3Dsoft-sph}). We also note
that the exponent $\alpha$ above gets larger for larger
system sizes for the MKA model. Although $N = 10000$ is a sufficiently
large system as far as any relevant length scale in this model is concerned, finite-size
effects as the orgin of the deviation from the AG relation in these
systems cannot be completely ruled out at present. 
Finally we note that the behaviour described here remains qualitatively 
the same if we use inverse diffusivities instead of $\alpha$-relaxation times.
Such a comparison will be presented elsewhere \cite{pap:Sengupta-etal-SEB}
in the context of the breakdown of the Stokes-Einstein relation.


In summary, we examine the validity of the  AG relation in different 
spatial dimensions by studying model liquids {\it via} computer simulations, for the 
first time in 2 and 4 dimensions.
The Adam-Gibbs relation is valid in four and three dimensions,
but is not obeyed in 2  dimensions and  the  nature of  the deviation from  the AG  
relation depends  on the details  of  the interaction  between  particles.
This is  an unexpected result. Although an understanding of this observation is lacking at
present, they present constraints that a successful theoretical
explanation of slow relaxation must meet. Such a lack of universality is unexpected and 
surprising since both Adam-Gibbs and RFOT theories describe the glass transition as a 
thermodynamic phase transition.

We thank SERC, IISc and CCMS, JNCASR for computational facilities. SSG acknowledges financial support from CSIR, India.


\begin{thebibliography}{}
\bibitem{pap:Ediger-review}M. D. Ediger, Annu. Rev. Phys. Chem. {\bf 51}, 99, (2000).
\bibitem{pap:angell-review-relaxation}C. A. Angell, K. L. Ngai, G. B. McKenna, P. F. McMillan, S. W. Martin, J. App. Phys., {\bf 88}, 3113 (2000).
\bibitem{pap:deb-stil}P. G. Debenedetti and F. H. Stillinger, Nature {\bf 410}, 259 (2001).
\bibitem{pap:AG-2}G. Adam and J. H. Gibbs, J. Chem. Phys. {\bf 43}, 139 (1965).



\bibitem{pap:RFOT-Kirkpatrick-etal}T. R. Kirkpatrick, D. Thirumalai, and P. G. Wolynes, Phys. Rev. A {\bf 40}, 1045 (1989).

\bibitem{parisi-mezard} G. Parisi and M. Mezard, {\it J. Chem. Phys.} {\bf 111} (1999).

\bibitem{mosaic} X. Xia and P. G. Wolynes, Proc. Natl. Acad. Sci. {\bf 97}, 2990 (2000)

\bibitem{rfotrev} V. Lubchenko and P. G. Wolynes, Annu. Rev. Phys. Chem. {\bf 58}, 235 (2007); 

\bibitem{pap:AG-RFOT-Bouchaud-Biroli} J. Bouchaud and G. Biroli, J. Chem. Phys. {\bf 121}, 7347 (2004).


\bibitem{pap:AG-Greet-Turnbull}R. J. Greet and D. Turnbull, J. Chem. Phys. {\bf 47}, 2185 (1967).
\bibitem{pap:AG-Richert-Angell-V}R. Richert and C. A. Angell, J. Chem. Phys. {\bf 108}, 9016 (1998).
\bibitem{pap:AG-Speedy} R. J. Speedy, Mol.Phys. {\bf 95}, 169 (1998).
\bibitem{pap:AG-Johari}G. P. Johari, J. Chem. Phys. {\bf 112}, 8958 (2000).
\bibitem{pap:AG-Scala-etal} A. Scala, F. W. Starr, E. La Nave, F. Sciortino and H. E. Stanley, Nature (London) {\bf 406}, 166 (2000).
\bibitem{pap:AG-Sastry}S. Sastry, Nature {\bf 409}, 164 (2001).
\bibitem{pap:AG-Mossa-etal}S. Mossa, E. La Nave, H. E. Stanley, C. Donati, F. Sciortino, and P. Tartaglia, Phys. Rev. E {\bf 65}, 041205 (2002).
\bibitem{pap:AG-Rolandetal}C. M. Roland, S. Capaccioli, M. Lucchesi and R. Casalini, J. Chem. Phys. {\bf 120}, 10640 (2004).



\bibitem{silvio_kac} S. Franz and A. Montanari  J. Phys. A: Math. Theor. {\bf 40}, F251 (2007).


\bibitem{dalle-ferrier} C. Dalle-Ferrier, C. Thibierge, C. Alba-Simionesco, L. Berthier, G. Biroli, J.-P. Bouchaud, F. Ladieu, 
D. L'H\^{o}te, and G. Tarjus, Phys. Rev. E  {\bf 76}, 041510 (2007).

\bibitem{pap:RFOTexpo-Capaccioli-etal} S. Capaccioli, G. Ruocco and F. Zamponi, J. Phys. Chem. B {\bf 112}, 10652 (2008).

\bibitem{birolietal-natphys} G. Biroli, J.-P. Bouchaud , A. Cavagna, T. S. Grigera and P. Verrocchio
{\it Nat. Phys.} {\bf 4}, 771 (2008). 


\bibitem{pap:RFOTexpo-Cammarota-etal} C. Cammarota, A. Cavagna, G. Gradenigo, T. S. Grigera, and
P. Verrocchio,  J. Chem. Phys. {\bf 131}, 194901 (2009).

\bibitem{pap:AG-Karmakar} S. Karmakar, C. Dasgupta and S. Sastry, Proc. Natl. Acad. Sci. (US),  {\bf 106},  3675 (2009).

\bibitem{thesis:Karmakar} S. Karmakar, Ph D. Thesis (2008) (web link: http://etd.ncsi.iisc.ernet.in/handle/2005/633.)


\bibitem{Shila} S. Sengupta, F. Vasconcelos, F. Affouard, and S. Sastry,  J. Chem. Phys. {\bf 135}, 194503 (2011).

\bibitem{pap:KA} W. Kob, H. C. Andersen, Phys. Rev. E {\bf 51}, 4626 (1995).

\bibitem{pap:KA-2D} R. Bruning , D. A. St-Onge1, S. Patterson and
Walter Kob, J. Phys.: Condens. Matter {\bf 21}, 035117 (2009).

\bibitem{pap:r10defnSK} S. Karmakar, A. Lema\^{i}tre, E. Lerner, and I. Procaccia,
 Phys. Rev. Lett. {\bf 104}, 215502 (2010).

\bibitem{pap:BC} D. Brown and J. H. R. Clarke, Mol. Phys. {\bf 51}, 5, 1243 (1984).


\bibitem{pap:4pt-CD}C. Dasgupta, A. V. Indrani, S. Ramaswamy and M. K. Phani, Europhys. Lett. {\bf 15}, 307 (1991).
\bibitem{pap:Ovlap-Glotzer-etal}S. C. Glotzer, V. N. Novikov and T. B. Schroder, J. Chem. Phys. {\bf 112}, 509 (2000).
\bibitem{pap:Ovlap-Donati-etal}C. Donati,  S. Franz, S. C. Glotzer, G. Parisi, J. Non-Cryst Solids {\bf 307}, 215–224 (2002).


\bibitem{pap:Sc-Sastry}S. Sastry,  Phys. Rev. Lett. {\bf 85}, 590 (2000).
\bibitem{pap:Sc-Sastry-JPCM}S. Sastry, J. Phys.: Condens. Matter {\bf 12}, 6515 (2000).
\bibitem{pap:PEL-Sciortino}F. Sciortino, J. Stat. Mech. P05015 (2005).
\bibitem{pap:PEL-Heuer}A. Heuer, J. Phys.: Condens. Matter {\bf 20}, 373101 (2008).


\bibitem{pap:sastry-deb-stil}S. Sastry, P. G. Debenedetti and F. H. Stillinger, Nature {\bf 393}, 554 (1998).

\bibitem{pap:onset-sastry} S. Sastry, PhysChemComm {\bf 3}, 79 (2000).


\bibitem{pap:atrep-karmakar} S. Karmakar, E. Lerner, I. Procaccia, Physica A, {\bf 391}, 1001 (2012).

\bibitem{pap:AG-Djugutovliq} Y. Gebremichael, M. Vogel, M. N. J. Bergroth, F. W. Starr and S. C. Glotzer, J. Phys. Chem. B {\bf 109}, 15068, (2005).

\bibitem{pap:AG-3Dsoft-sph} C. De Michele, F. Sciortino and A. Coniglio, J. Phys.: Condens. Matter {\bf 16}, L489 (2004). 

\bibitem{pap:Karmakar-PRL10}S. Karmakar, C. Dasgupta, and S. Sastry, Phys. Rev. Lett. {\bf 105}, 015701 (2010); 
S. Karmakar, C. Dasgupta, and S. Sastry, Phys. Rev. Lett. {\bf 105}, 019801 (2010).
\bibitem{pap:xiPTS-Hockyetal}G. M. Hocky, T. E. Markland and D. R. Reichman, {\it Phys. Rev. Lett.}, {\bf 108}, 225506 (2012).
\bibitem{pap:Sengupta-etal-SEB}S. Sengupta, S. Karmakar, C. Dasgupta and S. Sastry, (manuscript under preparation).
\end{thebibliography}
\end{document}